\documentclass[aps,prc,twocolumn,showpacs,superscriptaddress,groupedaddress]{revtex4-1}  

\usepackage{mathrsfs,amsmath,amssymb,amsthm,natbib}
\usepackage{amssymb,fmtcount}
\usepackage{graphicx,epsfig,latexsym,overpic,amssymb,color}
\usepackage{threeparttable}
\preprint{\today}

\begin{document}
\title{
Energy and pairing dependence of dissipation in real-time fission dynamics
}
\author{Yu Qiang}
\affiliation{
State Key Laboratory of Nuclear Physics and Technology, School of Physics,
Peking University, Beijing 100871, China
}
\author{J.C. Pei}\email{Corresponding author: peij@pku.edu.cn}
\affiliation{
State Key Laboratory of Nuclear Physics and Technology, School of Physics,
Peking University, Beijing 100871, China
}

\begin{abstract}

This work presents a microscopic study of dissipation  in fission dynamical evolutions with the time-dependent Hartree-Fock+BCS method in terms of energy dependence and pairing dependence. The friction coefficients and dissipation energies are extracted by mapping the symmetric fission process of  $^{258}$Fm  into a classical equation of motion.
Density-constrained  calculations are used to obtain the dynamical potential. The obtained friction coefficients have a strong dependence of deformations, and averagely match the coefficients adopted in statistical models.
The dissipation indeed increase with increasing initial excitation energies or with decreasing pairings.
It is also shown that post-scission dissipations play a significant role.  The demonstrated characteristics of dissipations will be valuable 
for calibrations of various fission models.

\end{abstract}
\maketitle

\section{Introduction}
For comprehensive and predictive descriptions of
correlated fission observables, a deeper understanding of fission mechanism is still strongly motivated~\cite{future}.
The dissipation and fluctuation are key ingredients in fission dynamics as well as in general non-equilibrium dynamical systems~\cite{Kubo_1966,Mavlitov,wenkai13,Simenel20}.
The dissipation effect refers to the exchange of energy, by all kind of damping mechanisms, from collective motion to intrinsic heat.
It was known that highly-excited nuclear fission involves strong viscosity, indicated by that experimental pre-scission neutron emission multiplicities are more than expected~\cite{statistical1}.
In semi-classical theory, the one-body dissipation can be depicted by the wall-and-window formula~\cite{BLOCKI}, which has been widely employed in modelings of nuclear dynamics~\cite{lede,wallwin1}.
The Langevin equation based on the over-damped assumption can well describe fission observables ~\cite{Randrup2}.
The dissipation is essential to describe fission evolution timescales in the later phase~\cite{Mavlitov} and corresponds to fluctuations through the Einstein relation~\cite{Kubo_1966}.
Strong dissipation is expected to result in long-neck fission configurations and strongly affect the total kinetic energies~\cite{koonin}, although distributions of fission yields are not sensitive to dissipation strengths~\cite{Sadhukhan_17}.
The amount of dissipated energies in fragments are key to describe the neutron multiplicities after fission and the spin population of fragments~\cite{randrup_21,nature}.
Usually the friction coefficients has been adjusted to reproduce experiments and the obtained coefficients are model dependent~\cite{Mccalla,WYe,krappe,shaw}.
Therefore microscopic studies of dissipation in fission is very desirable for validation of different transport models and statistical models.

From microscopic perspectives, the time-dependent density functional theory (TD-DFT) can naturally describe the non-adiabatic fission evolutions in the later phase of fission from saddle to scission~\cite{formation,Bulgacprl}.
TD-DFT is very successful in recent years  for describing the various post-fission observables, including fragment masses~\cite{stevenson1,stoprl,superfluid,me}, total kinetic energies (TKE)~\cite{stoprl,me}, excitation energies~\cite{Bulgacprc,me} and angular momentum of fragments~\cite{Bulgacangular}, which  agree reasonably with
experiments. TD-DFT is promising towards a unify microscopic framework for various correlated fission observables in the era with tremendous computing capabilities.
TD-DFT can in principle obtain collective variables such as the inertia of mass and dissipation coefficients, for connections with other models ~\cite{collectiveasp}.
TD-DFT only includes the one-body dissipation mechanism and the two-body dissipation can be investigated by beyond mean-field methods (e.g., time dependent density matrix method~\cite{TDDM}).
The one-body dissipation is dominant in low-energy fission processes and the two-body dissipation is expected to be important at higher energies~\cite{Tohyama,WenKai}.
However, the corresponding fluctuations are absent in TD-DFT.
There has been several approaches to build stochastic fluctuations into TD-DFT, aiming to obtain spreading fission observables ~\cite{stoprl,me}.
Therefore quantitative studies of dissipation are needed for calibrations of fluctuations in TD-DFT.
In nuclear collision and fusion reactions, a method called dissipative dynamics  has been proposed to extract the friction coefficients from TD-DFT to quantify the one-body dissipation effect~\cite{DDTDHF1,DDTDHF2,DDTDHF3},
in which the collective coordinates employ the relative distances and relative momentum between two nuclei.
This method has not been applied to fission studies yet.

In this work we employ the quadrupole deformation as the collective coordinate to extract the fission dissipation based on TD-DFT.
The dynamical potential is obtained by density-constrained calculations~\cite{densitycon},  which is similar to the work obtaining
nucleus-nucleus potential in fusion reactions~\cite{Umar}.
Therefore we can obtain deformation dependent dissipation coefficients from microscopic calculations.
Presently we study the dissipations from the symmetric fission of $^{258}$Fm as an illustrative example, corresponding to  one-dimensional classical equation of motion.
Furthermore, we aim to study the energy dependence and pairing dependence of dissipation coefficients.
Recently we carried TD-DFT calculations with initial wavefunctions from finite-temperature Hartree-Fock+BCS (TD-BCS) calculations~\cite{me}.
The resulted energy dependence of TKE and excitations of fragments are reasonable~\cite{me}.
The dissipations in highly excited fission process are expected to be significant and such energy dependence has
not been taken into account by the wall-and-window formula.
Thus the microscopic study of dissipations in compound nuclei with temperature or energy dependencies would be desirable.
In addition, pairing is crucial in fission dynamics to obtain reasonable fission observables~\cite{Bulgacprl,superfluid,me}.
Pairing can be seen as a lubricant in fission evolutions, in contrast to friction or dissipation.
The influences of pairings on dissipations would also  be of great interests.

\section{Theoretical framework}

Firstly the initial configurations of $^{258}$Fm are obtained by deformation-constrained Skrme-Hartree-Fock+BCS (HF-BCS) calculations with
different finite temperatures and pairing strengths.
The excitation energies of compound nuclei are related to the initial temperatures~\cite{FTHFB,pei}.
$^{258}$Fm has a typical symmetric fission mode and is ideal to calibrate fission dynamics~\cite{superfluid,stoprl,Bulgacsto}.
The constrained calculations are performed with SkyAX solver~\cite{SKYAX}, and all constrained deformations take as $(\beta_{2}, \beta_{3})=(1.65, 0.0)$.
Calculations in this work adopt the SkM$^{*}$ nuclear force ~\cite{skm*} and the mixed pairing interaction ~\cite{Mix}.
The pairing strengths take as $V_{p}$ =$-$480 MeV fm$^{3}$ and $V_{n}$ =$-$450 MeV fm$^{3}$ for protons and neurons, respectively.

The dynamical fission evolutions are investigated by the Sky3D solver ~\cite{sky3d} with the prepared initial configurations~\cite{Marko}.
The TD-BCS evolution is similar to the time-dependent Hartree-Fock-Bogoliubov (TD-HFB) formalism: $i\hbar \frac{\textrm{d} \mathcal{R}}{\textrm{d} t}= [H, \mathcal{R}]$ on the canonical basis ~\cite{CTDHFB}.
The initial HFB hamiltonian $H$ and the general density matrix $\mathcal{R}$ are associated with a finite temperature.
The box size ($x, y, z$) in TD-BCS is $48\times48\times64$ fm and the grid space is 0.8fm. The time step of evolution is 0.2 fm/c.

The macroscopic reduction procedure is realized by mapping the dynamical trajectory into a classical equation of the Langevin-equation form~\cite{DDTDHF2},
\begin{align}
    \dot{Q}_{} & = \frac{P_{}}{M_{}} \\
    \dot{P}_{} & = -\frac{\partial V_{dyn}}{\partial Q_{}}+\frac{1}{2}\frac{\partial M_{}}{\partial Q_{}}\dot{Q}_{}^{2}-\gamma(Q_{})P_{} \label{main}
\end{align}
where $Q_{}$ is the collective coordinate which adopts the quadrupole moment $Q_{20}=\langle 2\hat{z}^2-\hat{x}^2-\hat{y}^2 \rangle$; $P_{}$ is the corresponding collective momentum.
$\gamma (Q_{})$ is the friction coefficient, which is deformation dependent.  $M_{}$ denotes the quadrupole collective inertia of mass and $V_{dyn}$ is the dynamical nuclear potential.
The dissipation energy can be obtained by:
\begin{equation}
    E_{diss}(Q_{})=\int_{Q_{0}}^{Q_{}}\gamma (Q'_{}) P_{} dQ'_{} \label{Ediss}
\end{equation}
By mapping the microscopic dynamical evolution to the classical equation,
we can obtain the friction coefficient $\gamma(Q_{})$. In addition, the intrinsic energies in the evolution can
also be obtained by~\cite{lede},
\begin{equation}
    E_{int}(Q)=E^{*}-\frac{P^2}{2M}-V(Q, T=0),
\end{equation}
where $E^{*}$ is the initial excitation energy of the compound nuclei.

To obtain the dynamical nuclear potential $V_{dyn}$, we adopt the density constrained HF-BCS calculations.
This is different from the approach in Ref.~\cite{DDTDHF2} which employs two adjacent TDHF trajectories to get $V_{dyn}$ and $\gamma$ in fusion reactions.
In this work, density-constrained HF-BCS calculations are preformed to reproduce the density distributions from  TD-BCS calculations at each time step.
The density-constrained calculations have been adopted in studies of nuclear collisions in the literature ~\cite{Umar,Umar2}.

The collective inertia $M_{}$ can be derived by the method in ~\cite{collect}.
With the condition $ {\rm Tr}(\rho(t)[\hat{Q}_{}, \hat{P}_{}])=i\hbar$, $M_{}$  along the dynamical trajectory is obtained~\cite{collect}:
\begin{align}
    \frac{1}{M_{}} & =\frac{1}{m}{\rm Tr}[\rho \nabla Q_{}\cdot \nabla Q_{} ]
\end{align}
where $m$ denotes the nucleon mass.
Note that the collective inertia is derived non-adiabatically, which self-consistently includes dynamical effects and pairing effects.
Present work only solves the one-dimensional classical equation, aiming to study the major features of microscopic dissipations along elongations.
In principle, the collective friction and inertia of mass are tensor coefficients~\cite{lede}. The tensor inertia of mass can be obtained by
the multi-dimensional mapping method~\cite{collect}. However, the tensor friction coefficients are very difficult to be solved.

\section{Results}
\subsection{Energy dependence of dissipation effect}

It is known that the dissipation could be strong in hot nuclear matter~\cite{paul}. To describe the fission probability of highly excited compound nuclei, it is essential to
include a friction coefficient in the Kramers model to reproduce the experimental survival probability ~\cite{Mccalla, statdiss1, statdiss2}.
This is much concerned for synthesis of superheavy nuclei~\cite{survival}. The microscopic energy dependence of dissipation and friction coefficients are expected to provide valuable constraints on various
phenomenological models. There are several models for nuclear dissipation which  have very different energy dependence.
In Ref.~\cite{annal}, the derived two body dissipation from a collision dominance model decreases with temperature as $T^{-2}$.
However, the linear response theory predicts the dissipation increase with temperature ~\cite{hofman,yamaji}.
Besides, the nuclear dissipation from the macroscopic wall-and-window formula is nearly constant with increasing temperatures ~\cite{krappe}.

\begin{figure}[htbp]
\includegraphics[width=0.47\textwidth]{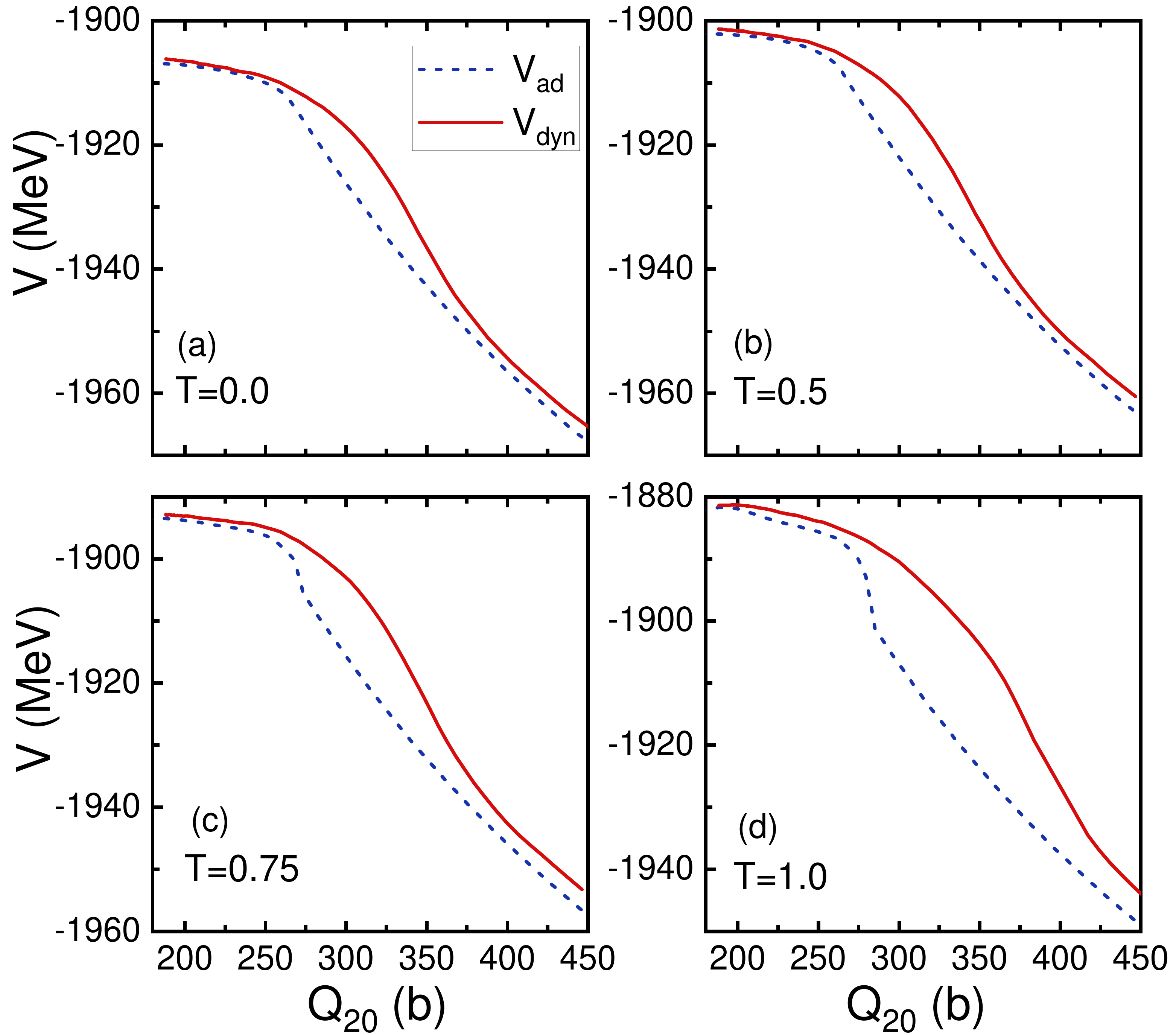}
\caption{\label{Fig11}
In the symmetric fission process of $^{258}$Fm, the nuclear potentials as a function of quadrupole potential $Q_{20}$ are obtained by static HF+BCS calculations (called adiabatic potential $V_{ad}$) and  density-constrained HF-BCS calculations with densities from TD-BCS evolutions (called dynamical potential $V_{dyn}$).
The potentials at different temperatures (or excitation energies) are shown: $T$=0 (a), $T$=0.5 MeV (b), $T$=0.75 MeV (c) and $T$=1.0 MeV (d).
}
\end{figure}

Firstly we study the dynamical nuclear potential in the fission process with comparison with the adiabatic potential, as shown in Fig.\ref{Fig11}.
Calculations at temperatures of 0.0, 0.5, 0.75 and 1.0 MeV are compared. The dynamical potential is obtained with density-constrained calculations ~\cite{densitycon} with densities from TD-BCS and
the adiabatic potential is obtained with deformation-constrained HF-BCS calculations. The fission process with temperatures above 1 MeV can not get scission
and stochastic fluctuations are needed~\cite{me}.
In Fig.\ref{Fig11}, it is shown that the dynamical potential and adiabatic potential are close at the beginning of the evolution until the turning point of the adiabatic potential.
Actually, the turning point or discontinuity is the scission point in the adiabatic fission.
The deformation of the turning point slightly increases with increasing temperatures.
There are large discrepancies between dynamical potential and adiabatic potential after the turning point.
Such discrepancies increase as temperature increases.
Of course, the adiabatic potential is always lower than the dynamical potential.
The dynamical potential is always smooth and the dynamical scission deformations are always larger than that of the adiabatic process.
The smaller adiabatic scission deformation indicates that adiabatic treatments would overestimate TKE.
After the dynamical scission, e.g., $Q_{20}$$\geqslant$400 $b$, the two potentials become close again as Coulomb potential dominates.

\begin{figure}[t]
\includegraphics[width=0.48\textwidth]{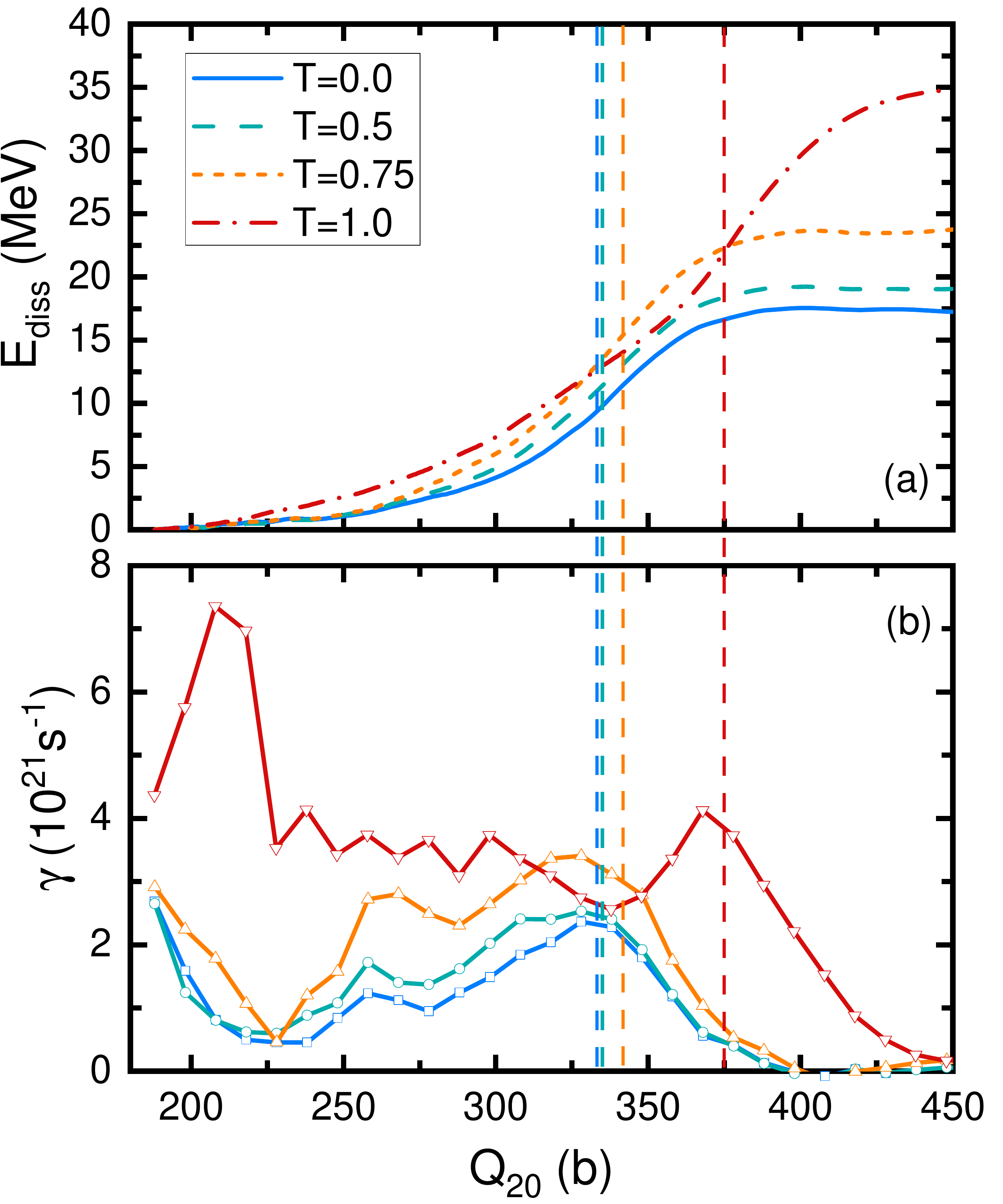}
\caption{\label{Fig22}
The panel (a) shows the dissipated energies obtained via Eq. \ref{Ediss} as a function of $Q_{20}$ at different temperatures.
The panel (b) shows the extracted friction coefficients $\gamma(Q)$ in Eq.\ref{main}. The vertical lines denote the scission deformations at different temperatures.
}
\end{figure}

After calculations of dynamical potentials, we can obtain the friction force $F_{fric}=-\gamma(Q_{})P_{}$
according to Eq.\ref{main}. The dissipated energy can be obtained by the integration in Eq.\ref{Ediss}.
The friction coefficient is derived by $F_{fric}$$/P_{}$. However,  fluctuations can appear in $\gamma(Q_{})$ in
numerical calculations especially in the beginning of evolution when the momentum is small. Thus
in this work, the friction coefficient is smoothed by averaging within a small deformation distance of 10 barn.

The derived friction coefficient and the dissipated energy as a function of quadrupole deformation $Q_{20}$ are shown in Fig.\ref{Fig22}.
It is shown that the dissipation is strongly dependent on nuclear shapes and fission evolutions.
The obtained friction coefficients averagely match the dissipation coefficients used in Langevin dynamical models and statistical models~\cite{krappe}.
For example, fission probabilities can be well reproduced by taking a friction coefficient of $3\times 10^{21} s^{-1}$~\cite{Mccalla}.
The original wall-and-window formula usually gives a very large dissipation parameter and a reduction factor has to be invoked ~\cite{krappe}.
Our results are consistent with the assumption that fission is very dissipative even at zero temperature.
We see that friction coefficients are considerable in the beginning of evolution.
With increasing deformations, the friction coefficients decrease while the fission process accelerates.
The dissipation of energies begins to accelerate due to increasing momentum $P$.
Then the dissipation increases when approaching the scission.
Finally the dissipation decreases after the scission while dissipated energies continuously increase until the friction becomes zero.
We can see that the dissipation of energy after scission is significant, which is related to the restoring process of the fragment shapes to equilibrium.
Therefore excitations of fragments should take into account both the before and after scission dissipations.

With increasing temperatures, the obtained friction coefficients indeed increase considerably.
The compound nuclei at temperature of 1 MeV ($E^{*}$=20.5 MeV) become highly dissipative with $\gamma$ around 4$\thicksim$6$\times 10^{21}s^{-1}$.
The dissipation energies also increase with increasing temperatures before scission.
On the other hand, the large dissipations at the beginning can hinder the fission process at high temperatures.
As we mentioned before that the TD-BCS calculations at temperatures above 1 MeV can not result in scission and fluctuation driving forces are essential~\cite{me}.
In addition, the scission deformations increases as temperatures increase. This is consistent with the knowledge that larger viscosity leads to longer necks~\cite{koonin}.

\begin{figure}[htb]
\begin{center}
\includegraphics[width=0.45\textwidth]{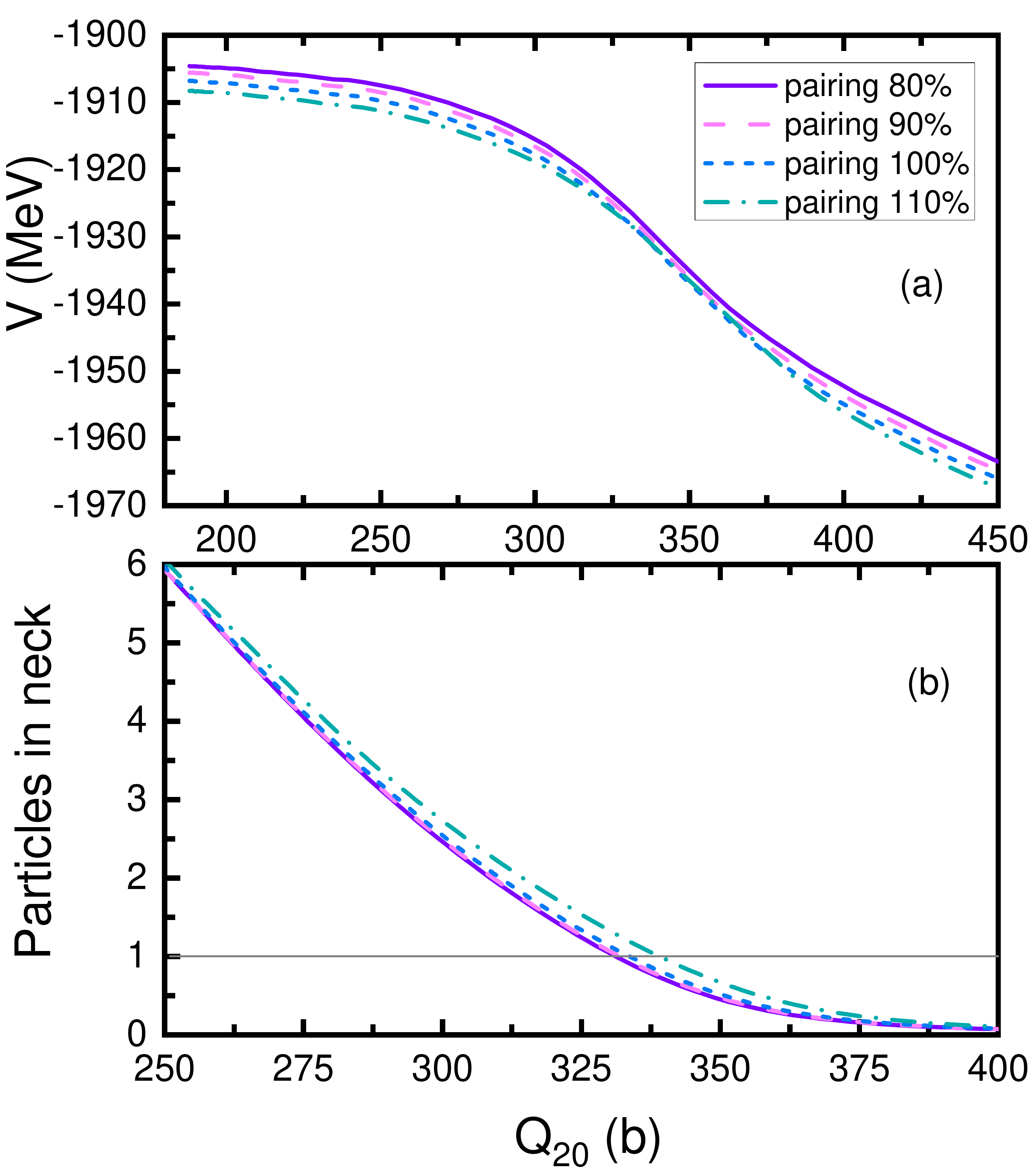}\\ 
\caption{\label{Fig33}
Panel (a) shows dynamical nuclear potentials as a function of $Q_{20}$ from density-constrained calculations with varying pairing strengths.
Panel (b) shows the number of particles in neck in the evolution, which is defined as the number of particles
in a neck of 2 fm length at the density minimum.
}
\end{center}
\end{figure}

\subsection{Pairing dependence of dissipation effect}

It is known that pairing plays an important role in fission dynamics~\cite{Bulgacprl,superfluid,me}.
In adiabatic treatments, pairing can accelerate the fission process by reducing fission barrier and inertia of mass~\cite{pair-s}.
In the fission dynamics, the dynamical pairing is essential to facilitate exchanges of levels via Landau-Zener effect~\cite{koonin}.
Pairing is often considered as a lubricant for the fission process, in contrast to the dissipation effects~\cite{future}.
For example, in some initial configurations,  TDHF without pairing can not evolved to scission while TD-BCS or TD-HFB can lead to scission~\cite{superfluid,me}.
In principle, the perfect superfluidity has no viscosity. It is of interests to study the pairing dependence of dissipations
in realistic fission dynamics.

\begin{figure}[t]
\includegraphics[width=0.45\textwidth]{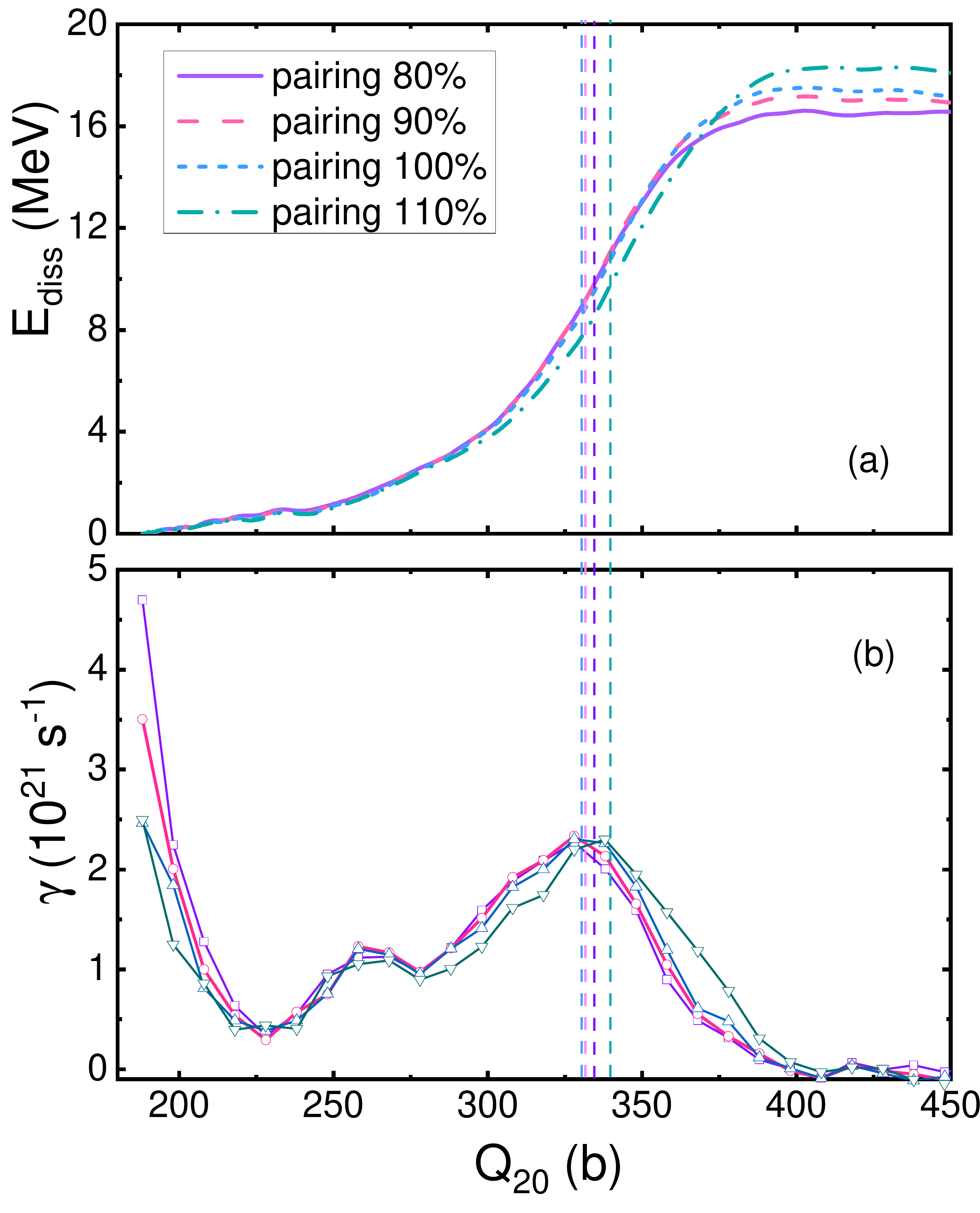}
\caption{\label{Fig44}
Similar to Fig.\ref{Fig22}, panel (a) shows the dissipated energy with varying pairing strengths from $80\%$ to $110\%$.
Panel (b) shows the obtained friction coefficients correspondingly. The scission deformations are denoted by the vertical lines.
}
\end{figure}

Fig.\ref{Fig33} displays the dynamical potential from density constrained calculations by varying the pairing strengths with a factor from $80\%$ to $110\%$.
Note that the same pairing is adopted for both HF+BCS as well as for TD-BCS calculations for self-consistency.
It can be seen that dynamical potentials don't differ much. It is reasonable that a larger pairing strength leads to a slightly lower potential.
Furthermore, the slope of potential, i.e. $\partial V_{dyn}/\partial Q$, changes around $Q_{20}$=345 $b$.
Note that a smaller slope of potential leads to a reduced friction force according to Eq.\ref{main}.
With increasing pairings, we see $\partial V_{dyn}/\partial Q$ becomes smaller when approaching $Q_{20}$=330 $b$ and then becomes larger when leaving $Q_{20}$=355 $b$.
This is consistent with the behaviors of friction coefficients in Fig.\ref{Fig44} from  $Q_{20}$=300 $b$ to 380 $b$ .
Fig.\ref{Fig33}(b) shows the number of particles in neck decreases slowly as a function of quadrupole deformations.
It can be seen that more particles in the neck with larger pairings.
It is expected that more particles in neck can enhance exchanges between single-particle levels.
Actually the number of particles in neck decreases rapidly towards scission in terms of evolution time~\cite{me}, although it decreases slowly in deformations.

\begin{table*}
  \caption{
 Calculated fission observables of  $^{258}$Fm including: the fission evolution time to scission $t_{s}$ (in fm/c),  scission deformations $Q_{20}$ (in $b$),  TKE (in MeV), the total excitation energies of fragments (TXE) (in MeV), pre-scission dissipated energies $E_{pre}^{diss}$ and total dissipated energies  $E_{tot}^{diss}$  (in MeV), the final intrinsic energies $E_{int}$ (in MeV).
 The results at different temperatures $T$ or excitation energies ($E^{*}$) are shown. The results with varying paring strengths by multiplying a factor $P_s$ from $80\%$ to $110\%$ are also shown. }
 \begin{tabular}{lccccccc}
  \hline
  \hline
 \hspace{15pt}  $T$ ($E^{*}$)  \hspace{15pt}  & \hspace{15pt} $t_{s}$  \hspace{15pt}    & \hspace{15pt}  $Q_{20}$  \hspace{15pt}   & \hspace{15pt} TXE  \hspace{15pt} & \hspace{15pt} TKE  \hspace{15pt}   &  \hspace{15pt} $E_{pre}^{diss}$  \hspace{15pt}  &  \hspace{15pt} $E_{tot}^{diss}$  \hspace{15pt}  & \hspace{15pt}  $E_{int}$  \hspace{15pt} \\
  \hline
 \multicolumn{3}{l}{Energy dependence} \\
0.0          & 804  & 334.1 & 29.7 & 239.8 & 9.6 &17.2 &17.04\\
0.5(4.7)   & 878  & 334.8 & 34.9 & 238.9 & 11.3 & 19.2  &18.21 \\
0.75(10.6) & 1080 & 342.2 & 45.9 & 235.8 & 15.5 & 23.9 & 21.59\\
1.0(20.5)  & 1732 & 374.7 & 64.5 & 226.7 & 21.9 & 35.1 & 30.35\\
  \hline
  \hline
 \hspace{2pt}  $P_{s}$ \hspace{2pt}  &  $t_{s}$   & \hspace{2pt}  $Q_{20}$  & TXE \hspace{2pt}  &   TKE  &  $E_{pre}^{diss}$ &  $E_{tot}^{diss}$ & $E_{int}$ \\
  \hline
\multicolumn{3}{l}{Pairing dependence} \\
80$\%$  & 1236 & 331.3 & 30.0   & 241.0 & 9.1 & 16.6 & 16.39 \\
90$\%$  & 904  & 331.7 & 29.7 & 240.6 & 9.2 & 17  &16.78 \\
100$\%$ & 804  & 334.1   & 29.7 & 239.8 & 9.6  & 17.2 &17.04\\
110$\%$ & 762  & 340   & 30.1 & 238.3 & 9.8 & 18.1 &17.91\\
  \hline
  \hline
\end{tabular}
  \label{tab1}
\end{table*}

Fig.\ref{Fig44} shows dissipated energies and friction coefficients with varying pairing strengths.
Generally differences in dissipation properties due to different pairings are not significant compared to that due to energy dependence.
We see that the friction coefficients are larger with decreasing pairing strengths at the beginning of the evolution.
Systems are very dissipative at the beginning with reduced pairings.
In addition, with larger pairings, the scission deformation increases.
The dissipated energies before scission are
close although friction coefficients are different.
This is understandable  that a larger pairing strength leads to larger momentum and smaller dissipation, and thus combined dissipated energies are close.
The final dissipated energies increase with increased pairings, and this is related to the dissipations after scission.
In fact, the larger scission deformations associated with large pairings contribute larger deformation energies for post-scission dissipations.
This is also true for the energy dependence of post-scission dissipations.
The reason for large scission deformations due to large pairings is different from that of the energy dependence, in which viscosity and scission deformation increase with temperatures.
In the case of increased pairings, the fission process is very lubricated with a large velocity, which just slides into large scission deformations.

\subsection{Other observables related to dissipation}

To have a comprehensive understanding of the dissipation effects, the evolution time,  scission deformations,  TKE, the total excitation energies of fragments (TXE), pre-scission and total dissipated energies, the final intrinsic energies are shown in
Table \ref{tab1}.
We see that the fission evolution time increases with increasing temperatures and decreases with increasing pairing strengths. This is consistent with the behaviors of the energy dependence of dissipation
and pairing dependence on dissipation.
TKE decreases with increasing temperatures and increasing pairings, which can be explained by the associated increased scission deformations.
Generally experimental TKE decreases with excitation energies~\cite{tke}.
It is known from TKE that $^{258}$Fm  has two symmetric fission modes with TKE around 205 and 230 MeV, respectively~\cite{Fm258tke}. Present TD-BCS result is close to the high-energy peak of TKE at 230 MeV, corresponding to the shorter-neck fission mode.
Note that TKE could be further reduced by including fluctuation effects~\cite{stoprl}.
TXE is approximately the sum of dissipated energy and the initial excitation energy.
We noticed that there is a discrepancy between TXE and  $E^{diss}_{tot}+E^{*}$, which could be ascribed to dissipations of other degrees of freedom beyond the quardpole deformation.
It can be seen that the post-scission dissipated energy plays a significant role.
With increasing pairings, the dissipated energies increase slightly although dissipation coefficients decrease.
This is because of the increased momentum $P$ in Eq.\ref{Ediss}, as indicated by the decreasing evolution times with increasing pairings.
Therefore, it is misleading to say that increased pairings lead to increased dissipations.
The intrinsic energies are usually been used to estimate  the intrinsic temperatures in Langevin equation~\cite{lede}. It is reasonable to see that intrinsic energies are close to obtained dissipated energies except differences appear at high temperatures.
With the evolving temperatures and dissipation coefficients, in principle we can determine the fluctuation amplitudes through the Einstein relation.
The two-body dissipation has not been included in this work and it could become important at high excitations.

\section{Conclusion}

We have presented the first microscopic study of energy dependence and pairing dependence of one-body dissipation effects in real-time fission dynamics. 
The friction coefficient is extracted by mapping the fission dynamics of $^{258}$Fm from TD-BCS into one-dimensional classical equation of motion.
The dynamical potential is obtained 
with density-constrained calculations.
The resulted microscopic friction coefficients are deformation dependent, ranging from 1 to 6 ($10^{21}s^{-1}$) and
averagely match the dissipations adopted in statistical models.  The fission is very dissipative even at zero temperature. 
This would be valuable for validations of other semi-classical fission models.
We demonstrated that friction coefficients increase with increasing temperatures or with decreasing pairing strengths.
Results show that fission at temperatures above 1 MeV is highly dissipative. 
Besides, it is shown that the post-scission dissipations play a significant role, which are associated with scission deformations.
The pairing as a lubricant accelerates the fission process as well as enhances exchanges between single-particle levels.
For better descriptions of fission observables, it is essential to include fluctuations in TD-DFT corresponding to dissipations.
The present study of dissipation behaviors will also be useful for calibrations of fluctuation strengths towards a predictive fission theory.

\acknowledgments
This work was supported by the National Natural Science Foundation of China under Grants
No.  11975032, 11835001, 11790325, and 11961141003,
an by the National Key R$\&$D Program of China (Contract No. 2018YFA0404403).
We also acknowledge that all computations in this work were performed in Tianhe-1A
supercomputer located in Tianjin.

\bibliographystyle{elsarticle-num_noURL}
\bibliography{elsarticle}

\end{document}